\documentclass[twocolumn,secnumarabic,amssymb, nobibnotes, aps, prl, superscriptaddress, nobalancelastpage]{revtex4-1}

\usepackage{graphicx}
\usepackage{amsmath}
\usepackage{epsfig}
\usepackage{graphicx}
\usepackage{stmaryrd}
\usepackage{amssymb,tabularx,dcolumn}
\usepackage{supertabular,ltxtable}
\usepackage{longtable}
\usepackage{float}
\usepackage{color}
\usepackage{bm}
\usepackage{CJKutf8}
\usepackage{hyperref}

\begin{document}

\begin{CJK*}{UTF8}{gbsn}

\title{First observation  of unbound $^{11}$O, the mirror of the halo nucleus $^{11}$Li}%

\author{T.B.~Webb}%
\email[Corresponding author: ]{tbwebb@go.wustl.edu}
\affiliation{Department of  Physics, Washington University, St. Louis, MO 63130, USA}
\author{S.M. Wang  (王思敏)}
\affiliation{National Superconducting Cyclotron Laboratory, Michigan State University, East Lansing, MI 48824, USA}

\author{K.W.~Brown}
\affiliation{National Superconducting Cyclotron Laboratory, Michigan State University, East Lansing, MI 48824, USA}
\author{R.J.~Charity}
\affiliation{Department of  Chemistry, Washington University, St. Louis, MO 63130, USA}
\author{J.M. Elson}
\affiliation{Department of  Chemistry, Washington University, St. Louis, MO 63130, USA}
\author{J.~Barney}
\affiliation{National Superconducting Cyclotron Laboratory, Michigan State University, East Lansing, MI 48824, USA}
\author{G.~Cerizza}
\affiliation{National Superconducting Cyclotron Laboratory, Michigan State University, East Lansing, MI 48824, USA}
\author{Z.~Chajecki}
\affiliation{Department of Physics, Western Michigan University, Kalamazoo, MI 49008, USA}
\author{J.~Estee}
\affiliation{National Superconducting Cyclotron Laboratory, Michigan State University, East Lansing, MI 48824, USA}
\author{D.E.M.~Hoff}
\affiliation{Department of  Chemistry, Washington University, St. Louis, MO 63130, USA}
\author{S.A.~Kuvin}
\affiliation{Department of Physics, University of Connecticut, Storrs, CT 06269, USA}
\author{W.G.~Lynch}
\affiliation{National Superconducting Cyclotron Laboratory, Michigan State University, East Lansing, MI 48824, USA}
\author{J.~Manfredi}
\affiliation{National Superconducting Cyclotron Laboratory, Michigan State University, East Lansing, MI 48824, USA}
\author{D.~McNeel}
\affiliation{Department of Physics, University of Connecticut, Storrs, CT 06269, USA}
\author{P.~Morfouace}
\affiliation{National Superconducting Cyclotron Laboratory, Michigan State University, East Lansing, MI 48824, USA}
\author{W. Nazarewicz}
\affiliation{Department of Physics and Astronomy and FRIB Laboratory, Michigan State University, East Lansing, MI 48824, USA}

\author{C.D.~Pruitt}
\affiliation{Department of Chemistry, Washington University, St. Louis, MO 63130, USA}
\author{C.~Santamaria}
\affiliation{National Superconducting Cyclotron Laboratory, Michigan State University, East Lansing, MI 48824, USA}
\author{J.~Smith}
\affiliation{Department of Physics, University of Connecticut, Storrs, CT 06269, USA}
\author{L.G.~Sobotka}

\affiliation{Department of  Physics, Washington University, St. Louis, MO 63130, USA}
\affiliation{Department of Chemistry, Washington University, St. Louis, MO 63130, USA}
\author{S.~Sweany}
\affiliation{National Superconducting Cyclotron Laboratory, Michigan State University, East Lansing, MI 48824, USA}
\author{C.Y.~Tsang}
\affiliation{National Superconducting Cyclotron Laboratory, Michigan State University, East Lansing, MI 48824, USA}
\author{M.B.~Tsang}
\affiliation{National Superconducting Cyclotron Laboratory, Michigan State University, East Lansing, MI 48824, USA}
\author{A.H.~Wuosmaa}
\affiliation{Department of Physics, University of Connecticut, Storrs, CT 06269, USA}
\author{Y.~Zhang}
\affiliation{National Superconducting Cyclotron Laboratory, Michigan State University, East Lansing, MI 48824, USA}
\author{K.~Zhu}
\affiliation{National Superconducting Cyclotron Laboratory, Michigan State University, East Lansing, MI 48824, USA}

\begin{abstract}
The structure of the extremely proton-rich nucleus $^{11}_{~8}$O$_3$, the mirror of the two-neutron halo nucleus $^{11}_{~3}$Li$_8$, has been studied experimentally for the first time. Following two-neutron knockout reactions with a $^{13}$O beam, the $^{11}$O decay products  were detected after two-proton emission and used to construct an invariant-mass spectrum. A broad peak of width 3.4~MeV was observed. Within the Gamow coupled-channel  approach, it was concluded that this peak is a multiplet with contributions from the four-lowest $^{11}$O resonant states: $J^{\pi}$=3/2$^-_1$, 3/2$^-_2$, 5/2$^+_1$, and 5/2$^+_2$.  The widths and configurations of these states show strong, non-monotonic dependencies on the depth of the $p$-$^9$C potential. This  unusual behavior is due to the presence of a broad threshold resonant state in $^{10}$N, which is  an  analog of the virtual state in $^{10}$Li in the presence of the Coulomb potential. After optimizing the model to the data,  only a moderate isospin asymmetry between ground states of $^{11}$O and $^{11}$Li was found.  

\end{abstract}

\date{present}%
\maketitle
\end{CJK*}

\textit{Introduction.--}
There is increasing interest in nuclei with large differences in their number of protons and neutrons as these can have unusual structures such as spatially extended halos and low-energy intruder states. One of the most iconic of these exotic nuclei is $^{11}$Li, which processes a large two-neutron halo that gives it a physical size similar to the much heavier $^{208}$Pb nucleus \cite{Tanihata:1996}. Borromean nuclei such as $^{11}$Li are  loosely bound three-body systems ($^9$Li core +2$n$) where there is no bound two-body subsystem \cite{Jensen2004}. In $^{11}$Li, the two valence neutrons have roughly similar probabilities of occupying the $(s_{1/2})^2$ and $(p_{1/2})^2$ configurations, with the former largely accounting for the halo \cite{Hagino2005}. The $n$+$^9$Li subsystem is also interesting as it may have an antibound, or virtual, state 
that could have important consequences for the halo structure of $^{11}$Li \cite{thompson1994,IdBetan:2004,Michel2006,SIMON2007267,Aksyutina2008}.

Mirror nuclei, with interchanged numbers of protons and neutrons, are expected to have similar nuclear structure due to isospin symmetry. In the case of $^{11}$Li, its  mirror partner  is the extremely proton-rich $^{11}$O nucleus  located beyond the proton drip line which has not been observed until this work. With 8 protons and only 3 neutrons, its nearest particle-bound neighbor is $^{13}$O, two neutrons away. Both valence protons  in $^{11}$O  are unbound making its ground state (g.s.) a two-proton (2$p$) emitter, similar to $^{12}$O \cite{Jager:2012}. 

The presence of  unbound nucleons presents an appreciable challenge for nuclear theory \cite{michel2009,michel2010a}.
The interaction between localized shell-model states and the continuum has been shown to lead to a number of interesting properties including clusterization \cite{okolowicz2012} and the breaking of mirror symmetry due to the Thomas-Ehrman effect \cite{grigorenko2004,michel2010}. There are predictions that the unbound $^{12}$O neighbor, has tens of percent more $(s_{1/2})^2$ occupation than its bound mirror $^{12}$Be \cite{grigorenko2004}.   Based on the extrapolation of the quadratic isobaric multiplet mass equation (IMME) fit to the  three neutron-rich members of the $A=11$ sextet,  the g.s. of $^{11}$O should be unbound by  3.21(84)~MeV  \cite{charityDIAS2012}, significantly more than $^{12}$O ($\sim$1.7~MeV). 
 Therefore, one might expect the  effect of the continuum coupling may lead to  even larger mirror symmetry breaking.  The presence of broad threshold resonant states in  the $p$+$^{9}$C scattering channel could  complicate these na\"ive expectations. 
Thus to understand the role of continuum couplings on the structure of $^{11}$O we 
have carried out both experimental and theoretical studies of this nucleus.

\textit{Experiment.--}
We produced $^{11,12}$O from one and  two-neutron knockout reactions with a $^{13}$O beam. Only a few results for $^{12}$O will be presented in this work.  The experiment was performed at the National Superconducting Cyclotron Laboratory at Michigan State University, which provided an $E/A=150$\,MeV $^{16}$O primary beam. This beam bombarded a $^{9}$Be target and the A1900 magnetic separator selected out $E/A$=69.5-MeV $^{13}$O fragments with a purity of only 10\%. The beam was then sent into an electromagnetic time-of-flight filter, the Radio Frequency Fragment Separator \cite{Bazin:2009}, and emerged with a purity of 80\%. This secondary beam impinged on a 1-mm-thick $^{9}$Be target and the charged reaction products  were detected 85\,cm further downstream  in the High Resolution Array (HiRA) \cite{WALLACE2007302} which consisted of 14 $\Delta$\textit{E-E} [Si-CsI(Tl)] telescopes. This array subtended  polar angles from $\theta_{lab.}$=2.1$^\circ$ to 12.4$^\circ$ in an arrangement almost identical to that in \cite{charity_carbon2011,brown2015,brown2017}. 

 Energy calibrations of the CsI(Tl) $E$ detectors were achieved with cocktail beams including $E/A=82.9$\,MeV $^9$C fragments and $E=80$\,MeV protons. Lower-energy calibration points were obtained using 2.2, 4.9, and 9.6\,mm-thick Al degraders. Monte Carlo simulations \cite{brown2015,brown2017} were used to determine the invariant-mass resolution. The accuracy of these simulations, and that of  the extracted centroid energies, were verified by studying the invariant mass of well-known levels, including the  $J^{\pi}$=2$^+_1$ and 2$^-_1$ excited states of $^{12}$N, which decay via the $p$+$^{11}$C channel  \cite{Jager:2012}.

\begin{figure}[htb]
\includegraphics[width=0.9\linewidth]{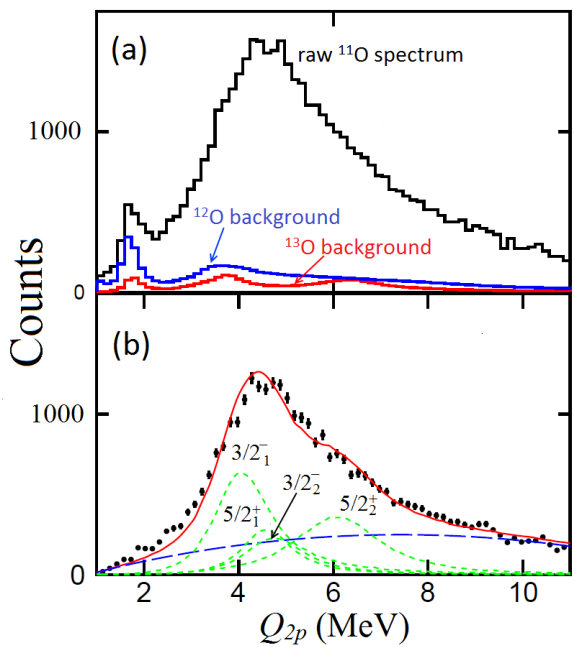}
\caption{Spectrum of total  energy $Q_{2p}$ released in the 2$p$ decay of $^{11}$O reconstructed from detected 2$p$+$^{9}$C events (a) including contamination from $^{10, 11}$C events and (b) with the contamination removed. The solid curve in (b) is a fit to the data composed of contributions from  four low-lying states (short dashed curves) predicted by theory while the long-dashed curve indicates the fitted background. To improve the experimental resolution, only events where $\lvert\cos(\theta)\rvert <$ 0.4 were included, where $\theta$ is the emission angle of the core in the $^{11}$O frame and $\theta$=0$^\circ$ is the beam axis \cite{Charity:2018}.}
\label{fig:O11composite}
\end{figure}

Figure \ref{fig:O11composite} shows the $^{11}$O total 2$p$-decay energy ($Q_{2p}$) spectrum reconstructed with the invariant-mass method from detected 2$p$+$^{9}$C events (a) before and (b) after background removal. This background results from fragments with undergo a nuclear reaction in a CsI(Tl) crystal, producing a smaller light output and are misidentified as a neighboring lighter isotope. From the calibration beams, we find 0.5\% of the $^{11}$C and $^{10}$C fragments were misidentified as $^9$C. The two background curves  in Fig.~\ref{fig:O11composite}(a) were determined from  2$p$+$^{10,11}$C coincidence events that were analyzed as if the detected carbon fragment was $^9$C.

 The background-subtracted spectrum displays a broad structure of peak energy $\sim$4.5 MeV, which is higher than the IMME extrapolation  for the g.s. of 3.21(84)~MeV. To estimate the width of this structure, we fit it assuming a diproton decay in the $R$-matrix prescription \cite{Barker:2001}, and incorporated the experimental resolution via the Monte Carlo simulations. However, we could not obtain a wide enough peak with the standard assumption for the diproton lineshape based on the $p$-$p$ phase shift.  This already casts doubt that the peak is a singlet. Instead to obtain larger widths, we took the lineshape as a delta function centered at  $E_{pp}$=0. The fitted intrinsic width was 3.4~MeV which is large compared to the experimental resolution (FWHM=0.45\,MeV at $Q_{2p}$=4.5\,MeV).

\textit{Theory.--}
To describe the spectra and two-proton ($2p$) decay of $^{11,12}$O, we utilized the three-body core+nucleon+nucleon Gamow coupled-channel (GCC) approach of Refs.~\cite{Wang2017,Wang2018} and we refer to these papers and to the review 
on the complex-energy shell model \cite{michel2009}
for technical details and basic concepts. The core ($^{9,10}$C) is chosen as a deformed rotor, which can reasonably reproduce the  intruder state and allows the pair of nucleons to couple to the collective states of the core. The wave function is constructed in Jacobi coordinates with a complex-energy basis, which can give exact asymptotic behavior of the wave functions and treats structure and reaction aspects on the same footing.

For the nuclear two-body interaction between valence nucleons, we took the finite-range Minnesota force~\cite{Thompson1977} augmented by their Coulomb interaction. The effective core-valence potential has  been taken in a deformed Woods-Saxon (WS) form including the spherical spin-orbit term \cite{Cwiok1987}. The Coulomb core-proton potential is calculated assuming the core charge is uniformly distributed inside its deformed nuclear surface ~\cite{Cwiok1987}.

 To analyze Thomas-Ehrman effects~\cite{Thomas1951,Ehrman1951,Thomas1952,Auerbach2000,michel2010}, the mirror nuclei with two valence neutrons, e.g., $^{11}$Li and $^{12}$Be, have been studied in a similar way. The deformed core is described by the quadrupole deformation $\beta_2$, and the couplings to the low-lying  rotational states are included. The core rotational  energies are taken from Ref.~\cite{ENSDF}. In the coupled-channel calculations, we included the g.s. band of the even-$A$ core with $J\le j_c^{\rm max} = 4^+$ and the odd-$A$ core with $J\le j_c^{\rm max} = 11/2^-$, respectively. According to  previous work~\cite{Wang2018}, the higher-lying rotational states have little influence on the final energy spectrum and g.s. mass.

Apart from the depth, the other parameters of the core-valence potential were optimized to fit the 1/2$_1^+$, 1/2$_1^-$, and 5/2$_1^+$ levels of $^{11}$N \cite{ENSDF} using the quasi-Newton method. At the $\chi^2$-minimum, the root-mean-square (rms) deviation is 197~keV.
The resulting values are: spin-orbit potential  $V_{\rm s.o.} = 15.09$\,MeV, diffuseness  $a=0.7$\,fm, the WS (and charge) radius $R=2.3$\,fm, and the quadrupole deformation $\beta_2 = 0.52$; they  are similar to those in Ref.~\cite{Fossez2016}, which can reasonably reproduce the intruder state of $^{11}$Be. The depth $V_{0}$ used in our $^{11}$O analysis, was  adjusted to fit the  1$_1^-$ and 2$_1^-$ states  of the core+nucleon system $^{10}$N with the rms error of 143~keV.

The GCC configurations can be described  both in the original Jacobi coordinates $(S,\ell_x,\ell_y)$  and the cluster orbital shell model coordinates $(j_1,j_2)$, where $S$ is the total spin of the valence nucleons and $\ell_x$ and $\ell_y$ are, respectively,  the orbital angular momenta of two-protons about their center of mass and of this center of mass about the core.
The calculations have been carried out in the model space of $\max(\ell_{x}, \ell_{y})\le 7$ with the maximal hyperspherical quantum number $K_{\rm max} = 20$. For the hyperradial part, we used the Berggren basis for the $K \le 6$ channels and the harmonic oscillator basis with the oscillator length $b = 1.75$\,fm and  $N_{\rm max} = 40$ for the higher-angular-momentum channels. The complex-momentum contour of the Berggren basis is defined as:  $k  = 0 \rightarrow 0.4-0.2i \rightarrow 0.6 \rightarrow 2 \rightarrow 4 \rightarrow 8$ (all in fm$^{-1}$), with each segment discretized with 60 points. To study antibound states and broad resonant states in the core-valence potential, we used the deformed  complex-momentum  contour as in Refs.~\cite{IdBetan:2004,Michel2006}.

\textit{Discussion.--}
To benchmark the theory, we calculated the $^{12}$O g.s. The experimental g.s. corresponds to a Breit-Wigner resonance having a centroid of $Q_{2p}$=1.688(29)\,MeV and an intrinsic width of $\Gamma$=51(19)\,keV. This experimental $Q_{2p}$ value is slightly larger than the  value of 1.638(24)\,MeV from a previous invariant-mass study~\cite{Jager:2012}, but still smaller than the values of 1.783(48), 1.760(24), and 1.740(22)\,MeV obtained from the $^{12}$C($\pi^+$,$\pi^-$) reactions \cite{fortune12O}.
Our calculations give $Q_{2p}$=1.97\,MeV and $\Gamma$=120\,keV, which is  in reasonable agreement with  experiment. If the depth $V_{0}$ of the WS potential is adjusted to reproduce the experimental $Q_{2p}$ value, the theoretical width is now 18$^{+4}_{-3}$\,keV, closer to our extracted value. The $(s_{1/2})^2$ configuration is predicted to account for 35\% of the strength compared to 20\% for the mirror nucleus $^{12}$Be calculated with the same parameters. This level of mirror symmetry violation is roughly in accord with the predictions of Ref.~\cite{grigorenko2004}. 

From the initial estimate of  $V_{0}$, we obtain  $Q_{2p}$=3.17\,MeV, $\Gamma$=0.86\,MeV for  $^{11}$O$_{g.s.}$,  which cannot explain the experimental peak.
To investigate whether this  peak  can be a singlet, we have varied the depth of the WS potential to the analyze how the decay width changes. If $Q_{2p}$ is set at  4.55\,MeV, i.e.,  close to the maximum of the experimental peak, the decay width is still only 1.29\,MeV with 67.7\% of ($K$, $S$, $\ell_x$) = (0, 0, 0) configuration. As this configuration is largely responsible for the decay rate \cite{grigorenko2004}, then we can obtain a maximum value of $\Gamma$=1.29/0.67$\approx$1.9\,MeV. Another rough estimate is obtained by assuming there is no valence-nucleon interaction and recoil term. Based on the three-body model, the g.s. energy and decay width of $^{11}$O would be two times as large as that of $^{11}$N. Hence, the decay energy $Q_{2p}$ should be around 4\,MeV with a width around 1.5\,MeV, which is still significantly less than  the experimental value. Thus, we conclude that the observed peak must contain multiple components.

We next consider the possibility that the experimental peak is a doublet  of 
the two predicted 3/2$^-$ states. Attempts to reproduce it by varying the depth of the WS potential to change $Q_{2p}$ and using the calculated line shapes to fit the data failed as the best-fit spectrum was still too narrow. Thus we are forced to also include  contributions from the two 5/2$^+$ states which would require the two knocked-out neutrons to come from different shells to conserve parity.  An analogous state has been observed in the mirror reaction, i.e., in two-proton knockout from $^{13}$B at a similar bombarding energy, the first-excited state of $^{11}$Li was observed \cite{Smith:2016} which has positive parity \cite{Kanungo2015,Tanaka2017}. This state is consistent with the 5/2$^+_1$ $^{11}$Li state obtained in our calculations.
The  energies  $Q_{2p}$ (and widths)  of the four lowest-lying resonant states in $^{11}$O obtained with $V_0$ optimized to the observed energy spectrum are:
4.16 (1.30)\,MeV for  $3/2^-_1$;
4.65 (1.06)\,MeV for  $5/2^+_1$;
4.85 (1.33)\,MeV for  $3/2^-_2$; and 
6.28 (1.96)\,MeV for  $5/2^+_2$. The statistical uncertainty in $V_0$ obtained from a $\chi^2$ analysis implies uncertainties of about 5\,keV on the centroids and 2\,keV on the widths. Based on our past experience, the centroids also have an additional $\pm$10~keV systematic uncertainty. 

The  $Q_{2p}$ value for the $3/2^-_1$  state is only 1.1$\sigma$ from the IMME extrapolation \cite{charityDIAS2012}.
  Figure~\ref{fig:O11composite}(b) shows the best fit with the contributions from these four  levels. 
 These line shapes have been modified to incorporate the experimental resolution. With  a smooth, rather flat background contribution included, the data are well described.
 The 3/2$^-_1$ and 5/2$^+_2$ peaks make up 39\% and 32\% of the fitted yield, respectively,  with the remainder attributed to  the 3/2$^-_2$ and 5/2$^+_1$
levels, which have similar centroids and widths and thus cannot be  disentangled.

\begin{figure}[!htb]
\includegraphics[width=0.8\linewidth]{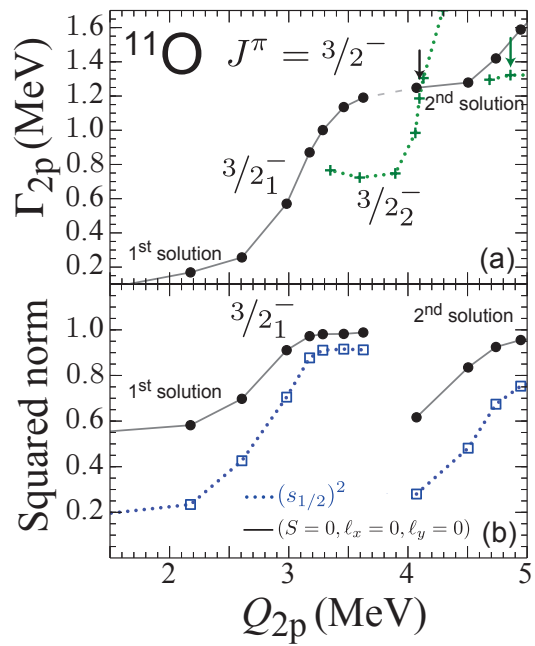}
\caption{\label{GSwidth} (a) predicted widths for the 
3/2$^{-}_1$ (both solutions) and 3/2$^{-}_{2}$ resonant states of $^{11}$O as a function of $Q_{2p}$. The arrows indicate the predicted $Q_{2p}$ values of the 3/2$^{-}_1$ and 3/2$^{-}_2$  states.
(b) GCC configuration weights (real parts of squared norms) of the $(s_{1/2})^2$
and $(S\ell_x\ell_y)=(000)$ configurations in the 3/2$^{-}_1$ wave function.
} 
\end{figure}

The dependence of the predicted widths  on the depth $V_{0}$, and hence $Q_{2p}$,  is complicated by the presence of broad  threshold resonant states in $^{10}$N, which affect the behavior of the $\ell=0$ single-particle channel.
Figure~\ref{GSwidth}(a) shows the $2p$ partial decay widths of the 3/2$^-$ resonant states of $^{11}$O as a function of $Q_{2p}$. The predicted  decay width of the 3/2$^-_1$ g.s. increases rapidly with $Q_{2p}$ below the Coulomb barrier ($\sim$3.2\,MeV); this is accompanied by a rapid configuration change, see Fig.~\ref{GSwidth}(b). The wave function of this state is dominated by the single $(S,\ell_x,\ell_y)=(000)$ Jacobi-coordinate component. For $Q_{2p}< 2.2$\,MeV, the  weight of the ($s_{1/2})^2$ shell-model component is around 20\%, which is similar to predictions for the mirror nucleus $^{11}$Li. As  $Q_{2p}$ gets larger, the  $(s_{1/2})^2$ amplitude rapidly increases. Close to the Coulomb barrier, this state becomes a pure $(s_{1/2})^2$ configuration. 
At energies above the barrier, the wave function  has a small amplitude inside the nuclear volume and can no longer be associated with an outgoing solution; it dissolves into the scattering continuum.  Interestingly, as seen in Fig.~\ref{GSwidth}(a), a second branch of the 3/2$^-_1$  solution  appears at higher $Q_{2p}$ values. This behavior is attributed to the presence of a broad resonant state in the $\ell = 0$ $p$-$^9$C scattering channel, which is an  analog of the antibound state of $^{10}$Li. When steadily increasing the Coulomb interaction from zero ($n+^{9}$Li) to the full $p+^{9}$C value at  $V_0 = -52.17$\,MeV ($Q_{2p} = 4.13$\,MeV), this resonant pole evolves in the complex-$k$ plane from the antibound state   in $^{10}$Li with  $k = -0.222i$\,fm$^{-1}$ ($E = -1.022$\,MeV), passing the region of
subthreshold resonances Re$(E)<0$ and $\Gamma>0$, located below the $-45^\circ$ line
in the momentum plane
\cite{Sofianos1997,Mukhamedzhanov2010}, and eventually becoming
the broad $s$-wave threshold resonant state in $^{10}$N with $k = 0.252-0.213i$\,fm$^{-1}$ 
($E = 0.38$\,MeV, $\Gamma=4.45$\,MeV), see Ref.~\cite{Efros1999,Lovas2002,Csoto2002,Mao2018} for more discussion. This antibound-state  analog is present in the broad range of $Q_{2p}$ values, and is the source of the discontinuity between the two branches for the 3/2$^-_1$ state when it gets close to the  $-45^\circ$ line.

The second 3/2$^-_1$ solution shows a similar trend to the first one, with the amplitude of the $(s_{1/2})^2$ configuration increasing with $Q_{2p}$. Our fitted value of $Q_{2p}$ for the g.s. corresponds to the second solution in Fig.~\ref{GSwidth} and contains 29\% of the $(s_{1/2})^2$ configuration compared a similar value of  25\% for the mirror nucleus $^{11}$Li. However at other values of $Q_{2p}$, the breaking of isospin symmetry would have been very substantial. 
The  3/2$^-_2$ state exhibits a similar  discontinuity as the 3/2$^-_1$ resonant solution; the 
dependence of the 3/2$^-_2$ $(s_{1/2})^2$ strength on $Q_{2p}$ is roughly inverted to that for  3/2$^-_1$ state in Fig.~\ref{GSwidth}(b), with excited-state  component dropping to zero when the ground-state approaches the pure $(s_{1/2})^2$ configuration.

\begin{figure}[!htb]
\includegraphics[width=1.\linewidth]{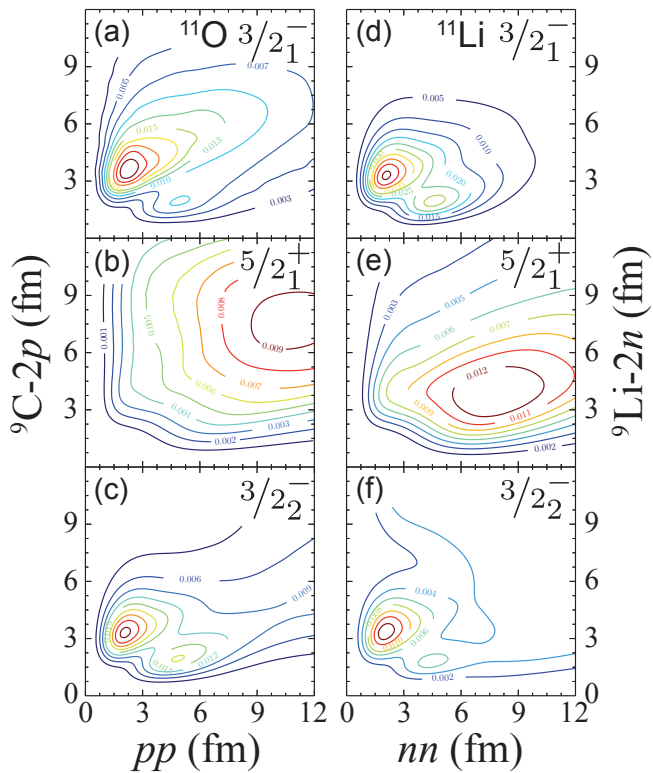}
\caption{\label{density} Two-nucleon density distributions (in fm$^{-2}$) in Jacobi coordinates predicted in GCC  for low-lying resonant states in $^{11}$O  (a-c)  and
 $^{11}$Li (d-f).}
\end{figure}

One can see the Thomas-Ehrman effect directly in the wave functions of the valence nucleons. Figure~\ref{density} shows the predicted two-nucleon density distributions in Jacobi coordinates (defined  in Ref.~\cite{Wang2017})  for the 3/2$^-_1$, 5/2$^+_1$, and 3/2$^-_2$ resonant states  of  $^{11}$Li and $^{11}$O obtained with our  optimized value of $V_{0}$ for the latter. The two 3/2$^-$ states  show strong correlations between the valence nucleons with either diproton or dineutron characteristics. As expected, the diproton peak in the unbound $^{11}$O is slightly less localized than that of the dineutron configuration in the bound mirror halo system. However, the  secondary peak strength for cigarlike arrangements is significantly reduced in $^{11}$O.
The two-nucleon correlations for the 5/2$^+_1$ state are rather weak, with the maximum density occurring for large separations between the two valence nucleons. 
Again, the 5/2$^+_1$ wave function is more extended spatially in $^{11}$O than in $^{11}$Li.
This resonant state is dominated by the $(s_{1/2},p_{1/2})$ component, while the 5/2$^+_2$ level in $^{11}$O is dominated by the $(s_{1/2},p_{3/2})$ component. 

\textit{Conclusions.--}
The proton-unstable isotope $^{11}$O has been observed for the first time. It was produced by two-neutron knockout reactions from a $^{13}$O beam. The invariant-mass spectrum of its 2$p$+$^9$C decay products measured with the HiRA detector contained a single broad peak with a width of  3.4\,MeV. The low-energy structure of $^{11}$O was also studied theoretically with the Gamow coupled-channel approach which suggests that the observed peak  is most likely a multiplet. We  obtained an excellent fit to this structure with contributions from the four-lowest excited resonant GCC states ($J^{\pi}$=3/2$^-_1$,5/2$^+_1$,3/2$^-_2$,5/2$^{+}_{2}$). 

The predicted width of the 3/2$^-_1$ g.s. shows complicated variation with the depth of the confining potential due to the presence of broad resonant states  in the $p$-$^9$C scattering $\ell=0$ channel. With our fitted  depth, the g.s.  configuration  was found to be similar to its mirror system $^{11}$Li. However, significantly different configurations were predicted for other values of the depth. These results demonstrate the importance of the coupling to the continuum for states beyond the drip lines and the role that $\ell=0$ near-threshold resonant states can play in constructing the many-body wave functions. We also studied 
the Thomas-Ehrman effect directly in the wave functions of valence nucleons in the mirror
$^{11}$O-$^{11}$Li pair. According to our calculations, the strength in the diproton (cigar) configuration is relatively stronger (weaker)  in the g.s. of $^{11}$O compared to  the situation in $^{11}$Li.

\textit{Acknowledgments.--}
This material is based upon work supported by the U.S. Department of Energy, Office of Science, Office of Nuclear Physics under award numbers DE-FG02-87ER-40316, DE-FG02-04ER-41320,DE-SC0014552, DE-SC0013365 (Michigan State University), DE-SC0018083 (NUCLEI SciDAC-4 collaboration), and DE-SC0009971 (CUSTIPEN: China-U.S. Theory Institute for Physics with Exotic Nuclei); and the National Science foundation under grant PHY-156556. J.M. was supported by a Department of Energy National Nuclear Security Administration Steward Science Graduate Fellowship under cooperative agreement number DE-NA0002135.

%

\end{document}